# Un méta-modèle pour l'automatisation du déploiement d'applications logicielles

**Noëlle MERLE**

*Equipe Adèle, laboratoire LSR – IMAG*
*Université Joseph Fourier*
*220, rue de la chimie*
*Domaine universitaire, BP 53*
*38041 Grenoble Cedex 9, France*
*Noelle.Merle@imag.fr*

RÉSUMÉ. Le déploiement est maintenant considéré comme une activité à part entière du cycle de vie du logiciel. Les grandes entreprises souhaitent pouvoir automatiser cette étape tout en prenant en compte les caractéristiques de chaque machine cible. Pour répondre à ces besoins, nous avons défini un environnement de déploiement : ORYA (Open enviRonment to deploY Applications). Cet environnement utilise un méta-modèle de déploiement, décrit dans ce papier. Notre approche utilise aussi les technologies des fédérations et des procédés, fournissant un environnement flexible et extensible pour l'utilisateur.

ABSTRACT. The deployment is now a full activity of the software lifecycle. Large enterprises want to automate this step, taking into account characteristics of each target machine. To satisfy these needs, we have defined an environment for the deployment phase: ORYA (Open enviRonment to deploY Applications). This environment uses a deployment metamodel, described in this paper. Our approach is based also on federation and process federations, providing a flexible and extensible environment to the user.

MOTS-CLÉS : déploiement, ORYA, méta-modèle de déploiement, environnement de déploiement.

**1. Introduction**

Le déploiement de logiciels a pris une grande importance avec le développement d'Internet et la complexité grandissante des logiciels. Il s'agit d'une étape complexe du cycle de vie du logiciel. Les entreprises s'intéressent aujourd'hui à l'automatisation de cette étape, tout en souhaitant prendre en compte les caractéristiques de chaque machine. Nous proposons une approche qui répond à ces besoins, en définissant un méta-modèle de déploiement, qui structure l'ensemble de l'information nécessaire au déploiement. Nous utilisons aussi les fonctionnalités de technologies existantes : fédérations et procédés. Les objectifs sont de fournir une approche adaptable, flexible et extensible pour l'utilisateur. La section 2 introduit succinctement le déploiement et ses concepts. La section 3 présente notre approche.

**2. Le déploiement**

Pour pouvoir fournir un support automatisé, il est nécessaire de prendre en compte les aspects d'organisation de l'entreprise pour définir les stratégies à utiliser, et, de considérer les caractéristiques logicielles et matérielles de chaque site cible.

Notre environnement définit une architecture à trois niveaux. Les applications sont développées au *niveau producteur*, puis acquises au *niveau entreprise*. L'entreprise peut ensuite ajouter ses propres extensions pour personnaliser l'application producteur. Le niveau entreprise permet aussi de préparer le déploiement physique vers les machines des utilisateurs, en prenant des décisions sur le choix des stratégies de déploiement à adopter. Puis, l'application est transférée et installée au *niveau utilisateur* (sur chaque machine cible).

Le déploiement arrive en fin de cycle de vie du logiciel. Cette phase se décompose elle-même en plusieurs activités qui constituent le cycle de vie du déploiement [MERL 04]. Celui-ci commence lorsqu'une nouvelle version d'une application est créée et installée sur un site client, et se poursuit jusqu'à sa désinstallation. Parmi les activités intermédiaires, on trouve la mise à jour et la reconfiguration, mais aussi l'activation ou la désactivation.

Tout au long de ce cycle de vie, l'intégrité des sites clients doit être conservée. Cette cohérence est définie par deux propriétés. La *propriété de réussite* permet d'assurer que l'application fonctionnera sur le site client, comme cela a été prévu et testé au niveau producteur (par exemple, on choisit la version à déployer en fonction des caractéristiques du site cible). La *propriété de sûreté* permet à une application déployée de ne pas détruire ou perturber, par effet de bord, les applications déjà installées. Le partage de composants (comme des DLL) entre applications peut provoquer de tels effets [PAR 01].

Il est important de fournir un environnement générique indépendant des applications à déployer, des outils de déploiement et de l'environnement de l'entreprise. Ainsi, tout type d'application peut être déployé en utilisant des outils différents et dans n'importe quelle entreprise. La section suivante décrit notre environnement de déploiement : ORYA (Open enviRonment to deploY Applications) qui tente de répondre à ces besoins.

## 3. ORYA

Notre travail s'inscrit dans le cadre du projet ITEA OSMOSE [OSM], au sein duquel notre équipe est responsable du groupe de travail sur le déploiement. En ce qui concerne l'état de l'art et les activités dans le domaine du déploiement le lecteur est invité à se référer aux travaux des différents partenaires de ce projet européen.

Notre environnement de déploiement, ORYA [LES 03], est basé sur la technologie des fédérations [EST 03]. Cette technologie, développée au sein de notre équipe, est une solution à l'interopérabilité et à la composition d'éléments hétérogènes. Une application est réalisée sous forme d'une fédération - dans notre cas, l'environnement de déploiement – obtenue en composant des éléments de haut niveau appartenant à des *domaines* distincts (domaine de procédés, domaine de produits, domaines métier, ....). La technologie des fédérations permet de coordonner les composants (outils) des différents domaines pour les faire travailler ensemble, et, de gérer les données communes contenues dans l'*univers commun*.

### 3.1. Un *méta-modèle conceptuel pour le déploiement*

Le méta-modèle de déploiement définit le domaine métier *déploiement* (décrit en UML par la Figure 1) et contient l'ensemble de l'information nécessaire pour le déploiement. Selon ce modèle, une entreprise (repère 1) est formée d'un ensemble de machines (3) qui sont structurées en groupes (2) et sous-groupes. Un groupe peut, par exemple, représenter le service informatique d'une entreprise, mais aussi un ensemble de machines avec des caractéristiques communes, comme le groupe des machines servant à la formation des personnels.

Une entreprise est aussi constituée d'un ensemble d'utilisateurs (4), qui représente les personnes pouvant se connecter au réseau de l'entreprise. Ces utilisateurs ont chacun au moins un rôle (5) (un administrateur, un directeur des ressources humaines, ...) et sont associés à la (les) machine(s) qu'ils utilisent.

Chaque machine est décrite par un ensemble de propriétés (6) et de contraintes (7). Par exemple, une propriété peut être le type de processeur ou de système d'exploitation de la machine ; une contrainte peut exprimer qu'une machine doit conserver un espace disque disponible supérieur à une certaine valeur. Le méta-modèle définit deux grandes catégories de machine : la catégorie "serveur d'applications" (8) et la catégorie "site client" (9).

Un serveur d'applications fournit un ensemble d'unités de déploiement qui sont packagées (10), c'est-à-dire que l'ensemble de l'information nécessaire pour le déploiement d'une unité est contenu dans une même entité.

L'information décrivant l'application (plus exactement le package) est représentée par un ensemble de propriétés (5) et contraintes (6). Cette information permet de sélectionner un package plutôt qu'un autre, en fonction de la configuration du site client et des souhaits de l'utilisateur. L'information de déploiement est représentée par un ensemble de ressources (11) et un procédé de déploiement (12). Les ressources sont notamment les fichiers nécessaires au cours du déploiement. Le procédé de déploiement décrit la liste des étapes (séquentielles ou parallèles) à réaliser pour déployer chaque unité du package.

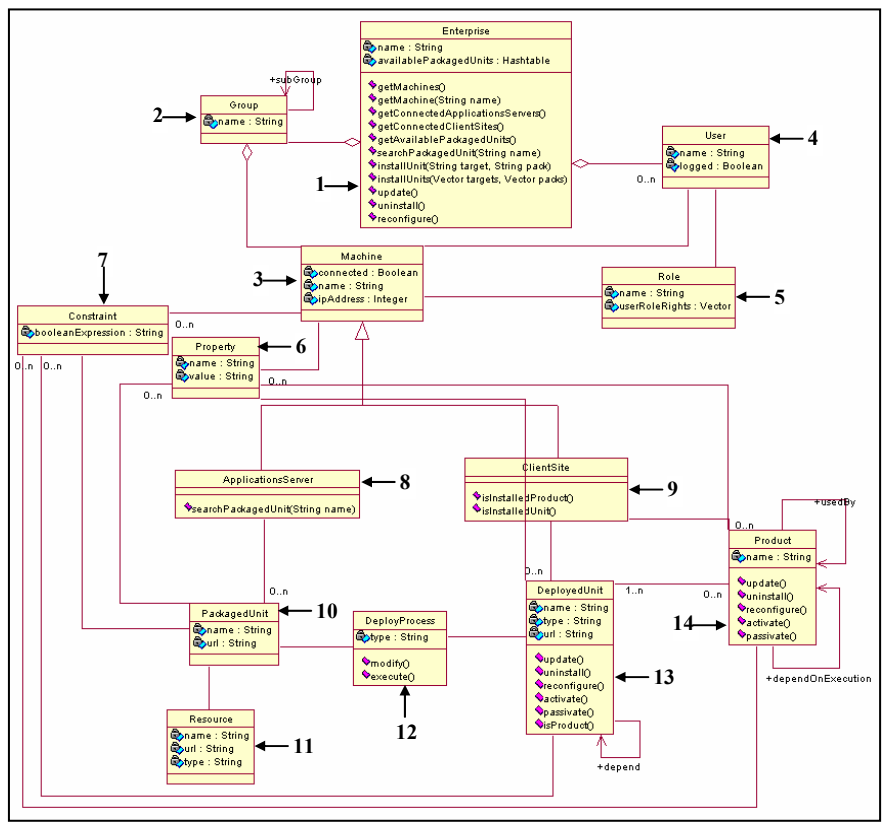

**Figure 1. Domaine de déploiement**

Comme nous utilisons la technologie des procédés [DER 99], le procédé de déploiement est interprétable par un moteur de workflow [WMC 95]. Un procédé décrivant comment atteindre un but et automatiser sa réalisation, le procédé de déploiement décrit comment déployer une unité. D'un point de vue fédération, l'exécution du procédé de déploiement est laissé à la charge du domaine de procédé (qui utilise dans notre cas le moteur de procédé de notre équipe : APEL).

D'autre part, un ensemble d'unités de déploiement (13) et de produits (ou applications) (14) sont déployés sur chaque site client (9). Un produit peut provenir de une ou plusieurs unités de déploiement (une unité contient tous les composants ou une partie des composants de l'application).

Un lien vers le procédé de déploiement utilisé pour déployer une unité est conservé. Il permettra d'effectuer une reconfiguration locale (dans le cas où des caractéristiques de la machine seraient modifiées) ou de demander une mise à jour en mode *pull* (demandée par l'utilisateur).

**3.2.** *Une fédération d'outils pour le déploiement*

Après avoir défini le méta-modèle qui constitue l'univers commun de notre fédération, il reste à définir et à affecter l'ensemble des rôles qui sont nécessaires pour réaliser le déploiement. Chaque rôle sera réalisé par un ou plusieurs outils (ou acteurs humains). Les outils ne sont utilisés que via les rôles qu'ils implémentent. Ainsi, chaque machine peut utiliser les outils dont elle dispose, dès qu'ils remplissent les rôles nécessaires.

Un premier rôle est défini comme le "service de déploiement" au niveau entreprise. Il s'agit d'avoir une vision globale de l'entreprise, et des groupes, machines et utilisateurs qui la composent. Il s'agit aussi de pouvoir décider d'une activité de déploiement en fonction des diverses stratégies de l'entreprise. Ceci permet de personnaliser le procédé de déploiement en fonction des besoins. Dans notre approche, ce service (rôle) est assuré par un "serveur de déploiement" (considéré comme un outil au sein de la fédération). Les déploiements en mode *push* sont lancés depuis ce serveur de déploiement. Il gère la communication entre les serveurs d'applications et les sites clients (qui ne se connaissent pas entre eux) et supervise le déroulement du déploiement.

Un rôle affecté aux serveurs d'applications permet d'ajouter ou de supprimer des unités packagées et d'obtenir les informations les concernant. Pour les sites clients, un autre rôle permet la gestion des contraintes et propriétés du site client. Les sites clients peuvent aussi consulter leur état (produits et unités déployées).

Les outils implémentant ces rôles utilisent les différents modèles génériques que nous avons définis. Ces modèles [MER 04] permettent de formaliser l'ensemble de l'information nécessaire au déploiement. Le *modèle de produit* (ou modèle d'application) décrit les caractéristiques des produits à déployer. Le *modèle de site* décrit les caractéristiques (logicielles et matérielles) des sites consommateurs (ressources et configuration) : il s'agit du contexte de déploiement. Le *modèle d'entreprise* décrit l'organisation interne de l'entreprise (machines, groupes de machines, utilisateurs,…). Enfin, le *modèle de procédé* décrit l'ensemble des activités à réaliser pour effectuer un déploiement (procédé de déploiement).

L'information contenue dans ces modèles permet de personnaliser et d'automatiser le procédé de déploiement selon les caractéristiques de chaque produit et de chaque site. De plus, un ensemble d'outils basiques de déploiement permet de réaliser chacune des étapes unitaires du procédé de déploiement (copie, transfert, chargement, activation, suppression, …).

## 4. Conclusion

Le déploiement de logiciels à grande échelle doit être automatisé tout en prenant en compte les caractéristiques de chaque machine et de chaque entreprise. L'utilisation d'un méta-modèle apporte une grande adaptabilité à notre approche. En effet, elle permet potentiellement de pouvoir déployer tout type de produits sur tout type de machine dans tout type d'organisation. Notre approche est aussi indépendante des outils de déploiement à utiliser, ce qui apporte plus de flexibilité pour l'utilisateur. Enfin, notre approche est supportée par les technologies des fédérations et des procédés, qui sont particulièrement bien adaptées à la problématique du déploiement.

## 5. Bibliographie

[DER 99] Derniame J.-C., Ali Kaba B., Wastell D., "Software Process : Principles, Methodology, and Technology", Springer-Verlag Berlin Heidelberg, 1999.

[EST 03] Estublier J., Le A.-T. and Villalobos J., "Multi-level Composition for Software Federations", *SC'2003*, Warsaw, Poland, April 2003.

[LES 03] Lestideau V., "Modèles et environnement pour configurer et déployer des systèmes logiciels", Thèse de l'université de Savoie, Décembre 2003

[MER 04] Merle N., Belkhatir N., "Une architecture conceptuelle pour le déploiement d'applications à grande échelle", *INFORSID 2004*, Biarritz, France, Mai 2004.

[MERL 04] Merle N., Belkhatir N., "Open Architecture for Building Large Scale Deployment Systems", *SERP'04*, Las Vagas, Nevada, USA, June 2004.

[OSM] OSMOSE, site internet du projet :  http://www.itea-osmose.org .

[PAR 01] Parrish A., Dixon B., Cordes D., "A conceptual foundation for component-based software deployment", *Journal of Systems and Software*, Volume 57, Issue 3, pages 193-200, 15 July 2001.

[WMC 95] The Workflow Management Coalition Specification, "The Workflow Reference Model", 1995, http://www.wfmc.org/standards/standards.htm.